\documentclass[aps,floatfix,twocolumn,footinbib,superscriptaddress,showpacs]{revtex4}
\usepackage{epsfig}
\usepackage{amsmath,amssymb}
\usepackage{bm}
\begin{document}

\title{Disorder effects on resonant tunneling transport in GaAs/(Ga,Mn)As heterostructures}


\author{Christian Ertler\footnote{email:christian.ertler@uni-graz.at}}

\author{Walter P\"otz}
\affiliation{Institute of Theoretical Physics, Karl-Franzens University Graz,
Universit\"atsplatz 5, 8010 Graz, Austria}


\begin{abstract}

Recent experiments on resonant tunneling structures comprising (Ga,Mn)As quantum wells
[Ohya et al., {\em Nature Physics} {\bf 7}, 342 (2011)] have evoked a strong debate regarding their interpretation 
as resonant tunneling features and the near absences of ferromagnetic order
observed in these structures. Here, we present a related theoretical study  of
a GaAs/(Ga,Mn)As double barrier structure based on a Green's function approach, studying the self-consistent interplay
between ferromagnetic order,
structural defects (disorder), and the hole tunnel current under conditions similar to those in experiment. 
We show that disorder has a strong influence on the
current-voltage characteristics in efficiently reducing or even
washing out negative differential conductance, offering an explanation for the experimental results.
We find that for the Be lead doping levels used in experiment the resulting spin density polarization in
the quantum well is too small to produce a sizable exchange splitting.

\end{abstract}

\maketitle

\section{Introduction}

Dilute magnetic semiconductors (DMS) are produced  by doping of
semiconductors with transition metal
elements, which provide local magnetic moments 
arising from open electronic $d$ or $f$ shells.\cite{Jungwirth2006:RMP,Burch2008:JMMM}
Bulk Ga$_{1-x}$Mn$_x$As may be regarded as the prototype:  Mn residing on the Ga site (Mn$_{\mathrm{Ga}}$) donates a hole, associated with valence band p-orbitals, and provides a local magnetic moment associated with partly filled Mn d-orbitals.   
 Mn$_{\mathrm{Ga}}$ is a moderately deep acceptor with the energy levels lying 
about 100 meV above the valence band edge.\cite{Schneider1987:PRL}  By increasing the Mn density the acceptor levels become more and more
broadened, developing into an impurity band which allows hole propagation and, for sufficiently high doping level, is believed to merge with the valence 
band.\cite{Jungwirth2007:PRB}  
At the same time, Mn more and more takes unwanted lattice positions, such as the antisite and interstitial position in the fcc lattice, or may even form Mn clusters, all leading to strong electron-hole compensation which eventually 
destroys  ferromagnetic ordering.   There is some debate as to the order in which these events occur as the Mn concentration is increased.  
Probably due to the presence of unintentional defects in (Ga,Mn)As samples, depending on growth
conditions, experimental evidence has led to 
somewhat conflicting conclusions about the precise position of the Fermi level in ferromagnetic bulk (Ga,Mn)As.\cite{Burch2008:JMMM} 
Some experiments can be interpreted by placing it into the 
top of a GaAs--like valence band edge which is broadened by disorder.\cite{Jungwirth2006:RMP}  Others suggest
the existence of an isolated impurity band in the ferromagnetic state.\cite{Richardella2010:S,Burch2006:PRL,Ohya2011:NP}

Recently a systematic series of experiments in form of non-equilibrium tunneling spectroscopy on double-barrier resonant tunneling structures with a (Ga,Mn)As quantum well 
\cite{Ohya2011:NP, Ohya2007:PRB, Ohya2010:PRL}
were conducted  to provide a deeper insight into this question. The group reported a near absence of ferromagnetic order in the well under bias and obtained
weak signatures of resonant tunneling, observable only in the second derivative of the current-voltage (IV) characteristic.  
Their conclusion that the Fermi energy lies in the impurity band has evoked strong debates and an alternative explanation has been given, 
which proposes that the resonant--tunneling signature is caused merely  by the
confined states in a potential pouch formed at the  contact/barrier heterointerface.\cite{Dietl:arXiv1102.3267D} In this explanation 
the observed dependence of the peak positions on the quantum well width is completely attributed to the increased series resistance which, however, 
seems to be insufficient to account for all well--width-dependent trends in the experimental results, as discussed in detail in a reply by Tanaka et al. 
which again emphasizes the existence of quantized levels in the (Ga,Mn)As quantum wells.\cite{Ohya:arXiv1102.4459O}

Indeed, quantization effects can be expected in (Ga,Mn)As for a layer thickness of about $\leq$ 3 nm since
in a recent scanning tunneling microscopy experiment the radius of the Mn acceptor wave function
has been determined  to be about 2~nm \cite{Richardella2010:S} and one can expect that near the band edge
Bloch-like and delocalized eigenstates will coexist in the picture of merging impurity and valence bands.\cite{Madelung:1978} 
Tunneling spectroscopic experiments of (Ga,Mn)As quantum well structures have
indicated such effects.\cite{Ohya2011:NP}  However, the signatures
in the current--voltage characteristics appear to be rather weak and
no regions of negative differential resistance due to resonances
associated with (Ga,Mn)As well layers have been observed as of yet, with the notable exception of an asymmetric magnetoresistance 
resonant--tunneling structure.\cite{Likovich2009:PRB}  This
suggests that a significant concentration of unwanted defects and/or disorder may be present, depending on growth conditions, as
it is known to be the case in thin layers of amorphous Si, in which
similarly weak signatures have been found.\cite{Miyazaki1987:PRL,Li1993:PRB} 
The density of imperfections due to the presence of Mn interstitial
or antisite defects  can be as high as 20\% of the nominal Mn doping, which makes (Ga,Mn)As  
a heavily compensated system.\cite{VanEsch1997:PRB,DasSarma2003:PRB}  Even lower structural quality for (Ga,Mn)As  
must be expected at heterointerfaces since the need of low-temperature epitaxy for growing the (Ga,Mn)As layers is harmful to forming clean interfaces with other materials. Moreover, the  interstitial defects may be trapped near the interfaces in post-growth annealing procedures which have been found successful 
for {\it bulk} (Ga,Mn)As. This suggests that transport through thin layers of (Ga,Mn)As is 
influenced by disorder and defects more severely than in annealed bulk structures.  

The growth of heterostructures, on the other hand,  provides  the appealing  opportunity  to drive (Ga,Mn)As layers into a genuine non-equilibrium situation by means of an external bias which
modifies their local hole density, possibly leading to  bias-dependent ferromagnetic behavior.\cite{Ertler2011:PRB, Ertler2012:JCE}  However, drawing conclusions 
from the physics of a thin (Ga,Mn)As layer
regarding the Fermi energy position in the bulk is an  intricate problem, since a Fermi 
energy in a (Ga,Mn)As quantum well under bias conditions is not well defined.   

 In a recent series of studies we have investigated 
the ferromagnetic bias anomaly in (Ga,Mn)As--based heterostructures and reached the conclusion that, for sufficiently high hole densities in the thin (Ga,Mn)As quantum wells,
 ferromagnetic ordering becomes bias dependent leading to variable spin-polarized currents.\cite{Ertler2011:PRB, Ertler2012:JCE}   Here, we study the low 
doping regime (relative to the Mn concentration) and use a refined model for the valence band states which accounts for both heavy and light--hole states.   
This allows us a direct comparison to recent experiments and, as will be shown, enhances the effect of disorder on suppressing 
a resonant--tunneling signature in the IV characteristic.
Based on a four band Kohn-Luttinger Hamiltonian the transport properties are investigated within a self--consistent  non-equilibrium Green's function method 
which accounts for space charge effects and a hole-density-dependent exchange splitting. 
We show that disorder reduces or even completely 
washes out regions of negative differential conductance in the IV curve.
We find that, for the Be lead doping levels as used in experiment, the resulting spin density polarization in
the quantum well is low and thus leads to  almost vanishing ferromagnetic order.  
Our theoretical model is presented in Sect.~\ref{sec:model} and the results and relevance to
experiment are discussed in Sect.~\ref{sec:results}.
Summary and conclusions are given in Sect.~\ref{sec:sum}.

\section{Physical Model}\label{sec:model}

Here we describe our transport model for heterostructures composed of layers of GaAs, GaAlAs, and (Ga,Mn)As grown along the z-axis.  
In this study the band structure of the top of the valence bands is modeled by the Kohn-Luttinger Hamiltonian \cite{Luttinger1955:PR}, which allows us to take into
account the 
mixing of heavy hole (HH) and light hole (LH) bands, which is of crucial importance for getting a realistic
transmission function for holes tunneling through a double-barrier structure as shown in Ref.~\onlinecite{Chao1991:PRB}. 
Ordering the four spin-3/2 basis vectors at the $\Gamma$-point as $u_\sigma = |\frac{3}{2},m_\sigma\rangle$ with
$m_\sigma = \{\frac{3}{2},\frac{1}{2},-\frac{1}{2},-\frac{3}{2}\}$, the wave-vector-dependent Kohn-Luttinger Hamiltonian reads 
\begin{equation}
 H(\mathbf{k}) = \left( \begin{array}{cccc}
             P+Q & -S & R & 0\\ -S^\dagger& P-Q& 0& R\\ R^\dagger & 0 & P-Q& S\\
0 & R^\dagger& S^\dagger& P+Q
            \end{array}
\right)~.
\end{equation}
The matrix elements can be expressed in terms of the dimensionless Luttinger parameters $\gamma_1,\gamma_2$ and $\gamma_3$:
\begin{eqnarray}
 P(\mathbf{k}) &=& \frac{\hbar^2}{2m}\gamma_1 k^2, \nonumber\\
 Q(\mathbf{k}) &=& \frac{\hbar^2}{2m}\gamma_2 (k_x^2+k_y^2-2k_z^2) \\
 S(\mathbf{k}) &=& \frac{\hbar^2}{2m}2\sqrt{3}\gamma_3 (k_x-i k_y)k_z, \nonumber\\
 R(\mathbf{k}) &=& \frac{\hbar^2}{2m}\sqrt{3}[-\gamma_2(k_x^2-k_y^2)+2i\gamma_3 k_x k_y]~, \nonumber
\end{eqnarray}
where $m$ is the free electron mass. In order to
considerably simplify the numerical demands for the calculation of macroscopic quantities, such as the current density, which require the summation over the
in-plane momentum, we apply the axial approximation in which the constant energy surface in the $\mathbf{k}$-space becomes cylindrically
symmetric but for which HH-LH band mixing is still included. Within the axial approximation the transmission function only depends on the absolute value of the in-plane momentum $k_\rho^2 = k_x^2+k_y^2$. Space-dependent
 (in $z$-direction) potentials are taken into account within the envelope function approximation, which effectively leads  to
replacing  $k_z\rightarrow$ by $-i\partial_z$.   By approximating the introduced
spatial derivatives on a finite grid of spacing $a$ one ends up with an effective nearest-neighbor tight-binding Hamiltonian 
of tridiagonal form 
\begin{equation}
H = \sum_{l,\sigma\sigma'} \varepsilon_{\sigma\sigma'}^{(l)} c_{l,\sigma}^\dagger c_{l,\sigma'}
+\sum_{l,\sigma\sigma'}t_{\sigma\sigma'} c_{l+1,\sigma}^\dagger c_{l,\sigma'} + \mathrm{h.c.},
\end{equation}
with $c_{l,\sigma}^\dagger$ denoting the creation operator for site $l$ and orbital $\sigma$. 
The on-site and hopping matrices, respectively,  take the form 
\begin{eqnarray}
\varepsilon_{\sigma\sigma'} &=& \left(
\begin{array}{cccc} 
C_1& 0 & -B & 0\\
0&C_2 & 0 & -B\\
-B& 0& C_2 &0\\
0&-B& 0&C_1
\end{array}\right),\\
\nonumber\\
t_{\sigma\sigma'} &=& \left(\begin{array}{cccc} D_1& -iE & 0 & 0\\
-iE&D_2 & 0 & 0\\
0& 0& D_2 &iE\\
0&0& iE&D_1           
\end{array}\right).
\end{eqnarray}
Here, the matrix elements are given by
\begin{eqnarray}
 C_1 &=& \frac{\hbar^2}{2m}[k_\rho^2 (\gamma_1+\gamma_2)+2(\gamma_1-2 \gamma_2)/a^2]\nonumber\\
  C_2 &=& \frac{\hbar^2}{2m}[k_\rho^2 (\gamma_1-\gamma_2)+2(\gamma_1+2\gamma_2)/a^2]\nonumber\\ 
B &=& \frac{\hbar^2}{2m}\sqrt{3}\overline{\gamma} k_\rho^2\\
 D_1 &=& \frac{\hbar^2}{2m}[-(\gamma_1-2\gamma_2)/a^2]\nonumber\\
      D_2 &=& \frac{\hbar^2}{2m}[-(\gamma_1+2\gamma_2)/a^2]\nonumber\\
 E &=& -\frac{\hbar^2}{2m}\gamma_3 k_\rho \sqrt{3}/a\nonumber
\end{eqnarray}
with $\overline{\gamma} = (\gamma_2+\gamma_3)/2$.
This effective tight-binding model has the advantage that space-dependent potentials, exchange splittings in magnetic layers, and 
  structural imperfections can be readily included in the orbital onsite energies of the model, i.e., the diagonal elements of
the onsite matrix using 
\begin{equation}\label{eq:DM}
\varepsilon_{\sigma\sigma}^{(l)} = {\varepsilon}_{\sigma\sigma}+ U_l - e \phi-\frac{\sigma}{2}\Delta_l +\varepsilon_\mathrm{rand}
\end{equation}
with $U_l$ denoting the intrinsic hole band profile due to the band
offset between different materials, 
$\phi$ is the the electrostatic potential, $e$ is the elementary charge, $\Delta_l$ denotes the local
exchange splitting in the magnetic materials with $\sigma =\pm 1$, and 
$\varepsilon_\mathrm{rand}$ labels a random shift due to disorder, as will be detailed below. 

With the ferromagnetic order being mediated by the itinerant carriers the
exchange splitting of the hole bands self-consistently depends on the local spin density of the holes. 
It can be derived within an effective mean-field model taking into account two correlated mean magnetic fields stemming
from the ions' d--electrons spin polarization $\langle S_z\rangle$ and the hole spin density 
$\langle s_z\rangle = (n_\uparrow-n_ \downarrow)/2$.\cite{Dietl1997:PRB,Jungwirth1999:PRB,Fabian2007:APS} 
The exchange splitting of the hole bands is then given
by
\begin{equation}\label{eq:delta}
 \Delta(z) = -J_\mathrm{pd} n_\mathrm{imp}(z) \langle S_z\rangle(z)~,
\end{equation}
with $z$ being the longitudinal (growth) direction of the structure,
$J_\mathrm{pd} > 0 $  is the exchange coupling  between the p-like holes
and the d-like impurity electrons, and $n_\mathrm{imp}(z)$ is
the impurity density profile of magnetically active ions. The
effective impurity spin polarization $\langle S_z\rangle$ is induced by the magnetic field caused by the mean hole spin
polarization, yielding
\begin{equation}\label{eq:Szgen}
  \langle S_z\rangle= - S B_S\left( \frac{S J_\mathrm{pd} \langle s_z \rangle}{k_B T}\right),
\end{equation}
where, $k_B$ is the  Boltzmann constant,  $T$ is the lattice
temperature, and $B_S$ is the Brillouin function of order $S$, here with
$S = 5/2 $ for the Mn impurity spin. Combining the last two expressions gives the desired result
\begin{equation}\label{eq:delta1}
 \Delta(z) = J_\mathrm{pd} n_\mathrm{imp}(z)
S B_S\left\{\frac{S J_\mathrm{pd} [n_\uparrow(z)-
n_\downarrow(z)]}{2 k_B T}\right\}.
\end{equation}
Since the hole spin density $\langle s_z\rangle$ is changed by the in- and out-tunneling holes, the magnetic and transport properties
of the system are coupled self-consistently.

To obtain realistic potential drops between the two leads 
space-charge effects have to be taken into account.
In the Hartree approximation the electric potential is determined by the Poisson equation,
\begin{equation}\label{eq:poisson}
 \frac{\mathrm{d}}{\mathrm{d}z} \epsilon \frac{\mathrm{d}}{\mathrm{d}z}\phi =
e\left[ N_a(z) - n(z)\right],
\end{equation}
where $\epsilon$ and $N_a$, respectively, denote the dielectric constant and the
acceptor density. The
local hole density at site $|l\rangle$ can be obtained from the non-equilibrium ``lesser''
Green's function $G^<$:
\begin{equation}\label{eq:n}
 n(l) = \frac{-i}{A a}\sum_{\mathbf{k}_{\|},\sigma}\int\frac{\mathrm{d}E}{2\pi} G^<(E;l\sigma,l\sigma)~,
\end{equation}
with $A$  and $\mathbf{k}_{\|}$, respectively, being the in-plane cross sectional area of the structure and the in-plane momentum. 
 The lesser Green's function is determined by the equation of motion
\begin{equation}\label{eq:gless}
 G^< = G^R\Sigma^<G^A
\end{equation}
where $G^R$ and $G^A = [G^R]^+$ denotes the retarded and advanced Green's function, respectively.
The scattering function $\Sigma^<=\Sigma^<_l+\Sigma^<_r$ describes
the particle inflow from the left $(l)$ and right $(r)$ reservoirs  \cite{Datta:1995} with
\begin{equation}
\Sigma^<_{l,r} = f_0(E-\mu_{l,r})(\Sigma^A_{l,r}-\Sigma^R_{l,r})~,
\end{equation}
where $f_0(x) = [1+\exp(x/k_B T)]^{-1}$ is the Fermi distribution function and
$\mu_l$ and $\mu_r$, respectively, denote the quasi--Fermi energies in the contacts.
The retarded and advanced self-energy terms $\Sigma^R = \Sigma_l^R+\Sigma_r^R$ and $\Sigma^A = [\Sigma^R]^+$, respectively, 
couple the simulated system region to the left and right contacts.  The surface Green's function of the leads is needed to
obtain the contact self-energy $\Sigma^R$  
and is calculated by using the algorithm of L{\'o}pez-Sancho et al.\cite{Sancho1985:JPF}
The retarded Green's function of the system, finally, is given by
\begin{equation}\label{eq:gr}
 G^R = \left[E+i\eta-H_s-\Sigma^R\right]^{-1}~.
\end{equation}
which we calculate by consecutively adding one layer of the system after another which, in our case,  solely requires the inversion
of a 4x4 matrix for each additional layer.

The transport equations Eqs.~(\ref{eq:gless}) and (\ref{eq:gr}) couple via the
spin-resolved hole density to the exchange splitting of the hole bands Eq.~(\ref{eq:delta1}), and the
Poisson equation Eq.~(\ref{eq:poisson}). For a given applied voltage this system of equations is
solved in a self-consistent loop until convergence of the electrostatic potential and the exchange field is reached.
A small external magnetic field is applied initially to aid spontaneous symmetry breaking.  
For the next bias iteration, the self--consistent solution from the previous bias value is used for an initial guess.  
Having obtained the self-consistent potential profile
the transmission probability $T(E,\mathbf{k}_{\|})$
from the left to the right reservoir is calculated
by 
\begin{equation}
T(E,\mathbf{k}_{\|}) = \mathrm{Tr}\left[\Gamma_l G^R \Gamma_r G^A\right]
\end{equation}
with $\mathrm{Tr}[\cdot]$ denoting the trace operation and 
the lead coupling functions being defined by $\Gamma_{(l,r)} = i[\Sigma_{(l,r)}^R-\Sigma_{(l,r)}^A]$.

The steady--state current density is obtained by an integration over all incoming $\mathbf{k}_{\|}$ states and the total energy $E$:
\begin{equation}
j = \frac{e}{2\pi\hbar}\sum_{\mathbf{k}_{\|}} \int \mathrm{d} E\: T(E,\mathbf{k}_{\|})\left[f_l(E,\mu_l) -f_r(E,\mu_r)\right]
\end{equation}
with the applied bias $V=(\mu_l-\mu_r)/e$  being defined as the difference
in quasi-Fermi levels of the contacts.

We model disorder effects in the (Ga,Mn)As layers by performing a configurational average over structures with randomly
chosen diagonal elements of the onsite matrix resulting in an ensemble of tight-binding Hamiltonians. 
The diagonal onsite matrix elements are sampled according to a Gaussian distribution around $\varepsilon_\mathrm{rand}=0$ in Eq.~(\ref{eq:DM}) for increasing standard deviations 
($\sigma_\mathrm{dis} =$ 20, 40, 60, and 80 meV).
For each specific structure (and Hamiltonian) the current-voltage (IV) curve is solved
self--consistently for an upsweep of the applied voltage. 
This effective one--dimensional modeling of disorder must be viewed as a limited estimate since it corresponds to a  
cross--sectional average of  transport though 
uncorrelated effective linear chains.  As such, any disorder correlations parallel to the interface are neglected.  
Such correlations in disorder will play a role in the establishing of ferromagnetic order in real structures relative to the idealized 
homogeneous mean--field model adopted here, since both ferromagnetic order and disorder effects are highly dependent upon spatial
 dimensionality.\cite{Kaxiras:2003, Ashcroft:1976}   Also, this type of averaging cannot model the effects of Mn clusters.
Additional types of disorder from Mn clustering, Mn interstitials, etc. however may be present in real structures, but their type and concentration may differ from sample to sample.


\section{\label{sec:results} Results}

We investigate a double barrier structure consisting of Be-doped GaAs leads and a (Ga,Mn)As quantum well,
similar to the experimental setups presented in Refs.\onlinecite{Ohya2007:PRB,Ohya2010:PRL}. 
The structural similarity of the layer materials allows one to use the same valence band model for the whole structure, which considerably
simplifies the theoretical description.
In recent resonant tunneling spectroscopic experiments of the group of Tanaka \cite{Ohya2011:NP}, however,  a Schottky barrier contact with Au at one side and a GaAs:Be contact with an AlAs barrier on the other side of the (Ga,Mn)As quantum well
was used. If the
transport is primarily determined by the resonant levels in the well, our results obtained for a completely zincblende structure will be relevant also for
the Schottky-barrier system investigated in experiment.

For the simulations we assume following parameters, comparable to the experimental values of Ref.~\onlinecite{Ohya2007:PRB}:
barrier thickness $d_\mathrm{bar} = 4$ layers ($\approx$ 2 nm), quantum well width $d_\mathrm{w} = 10$ layers ($\approx 5$ nm), barrier height $V_\mathrm{bar} = 300$ meV,
relative permittivity of GaAs $\epsilon_r = 12.9 $, exchange coupling constant $ J_{\mathrm{pd}}= 0.06 $~eV nm $^3$, $5\%$ Mn doping, and
the lead temperature  $T = 4.2$~K. The Fermi level in the GaAs leads is chosen to $\mu_l = 10$~meV, which corresponds to a Be-doping of about $10^{18}$cm$^{-3}$
as used in experiment. The dimensionless Luttinger parameters are set to standard values for GaAs: $\gamma_1 = 6.85, \gamma_2 = 2.1$, and $\gamma_3 = 2.9$.

\begin{figure}[!t]
\centering
\includegraphics[width=0.95\linewidth]{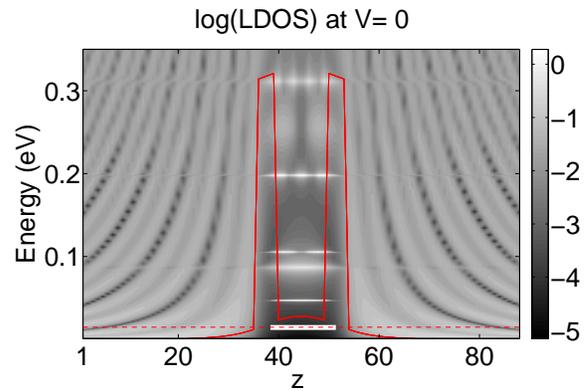}
\caption{(Color online) Logarithm of the local density of states (LDOS) at zero bias for the perfect structure (no disorder) using
an inverted energy scale. 
The self--consistent potential profile is indicated by the (red) solid line, whereas the
Fermi energy position of the leads is plotted as (red) dashed  line.
The position of the impurity band in the well is indicated schematically  by the (white) bold solid line below the valence band edge.}
\label{fig:ldos1}
\end{figure}

\begin{figure}[!t]
\centering
\includegraphics[width=0.95\linewidth]{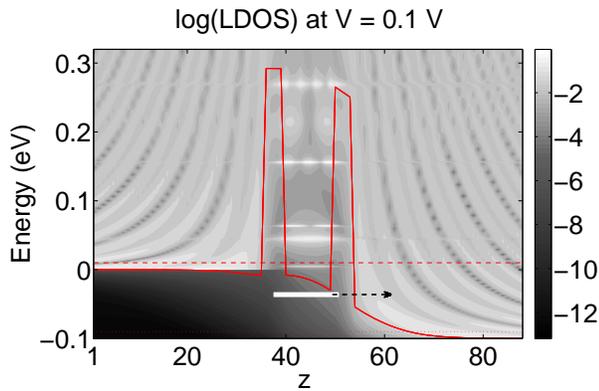}
\caption{(Color online) Logarithm of the local density of states (LDOS) at the bias $V=0.1$~V for the perfect structure (no disorder) using
an inverted energy scale. 
The self--consistent potential profile is indicated by the (red) solid line. The
Fermi energy position of the left and right lead is plotted as (red) dashed and dotted line, respectively.
Holes of the impurity band (indicated schematically by the bold (white) solid line below the valence band edge) can tunnel out to the collector
reservoir.}
\label{fig:ldos2}
\end{figure}

\begin{figure}[!t]
\centering
\includegraphics[width=0.95\linewidth]{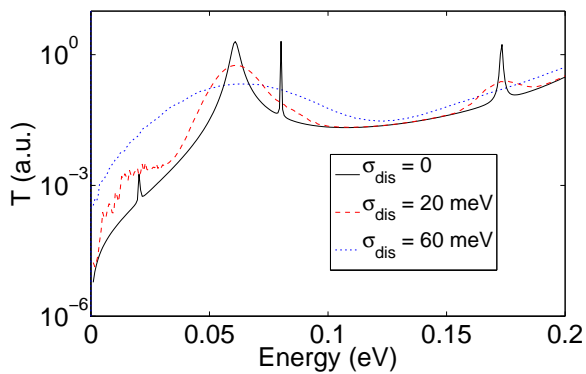}
\caption{(Color online) The transmission function at zero bias for the perfect structure (solid line),
for moderate disorder of $\sigma_\mathrm{dis} = 20$~meV (dashed line), and stronger disorder $\sigma_\mathrm{dis} = 60$~meV (dotted line) by sampling
over $3\times10^4$ configurations.}
\label{fig:T}
\end{figure}

\begin{figure}[!t]
\centering
\includegraphics[width=0.95\linewidth]{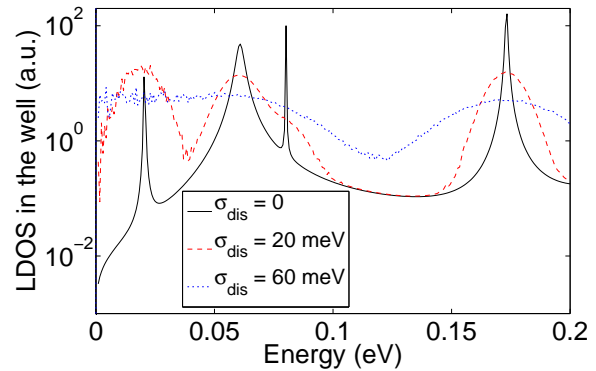}
\caption{(Color online) Local density of states (LDOS) in the middle of the quantum well at zero bias for the perfect structure (solid line),
for moderate disorder of $\sigma_\mathrm{dis} = 20$~meV (dashed line), and stronger disorder $\sigma_\mathrm{dis} = 60$~meV (dotted line) by sampling
over $3\times10^4$ configurations.}
\label{fig:wdos}
\end{figure}

\begin{figure}[!t]
\centering
\includegraphics[width=\linewidth]{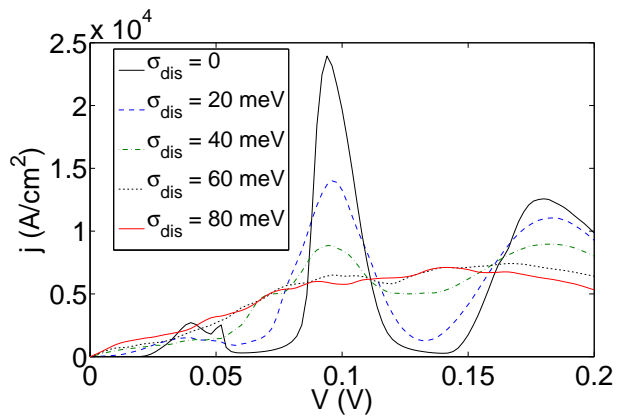}
\caption{(Color online) IV-characteristic of a magnetic double barrier structure for increasing degree of 
disorder measured by the standard deviation $\sigma_\mathrm{dis}$ of the Gaussian distribution of the randomly
chosen diagonal onsite matrix elements $\varepsilon_\mathrm{rand}$ .}
\label{fig:IVkl}
\end{figure}

While (Ga,Mn)As inevitably is a disordered system, our starting point is an idealized system for which the valence band edge is identical to that of GaAs.  
Subsequently, valence band disorder is added in increasing steps allowing for a systematic qualitative account of its consequences on the IV-curve.  
For convenience  we use an inverted hole energy scale but retain positively charged holes.
In order to simulate the presence of Mn$_\mathrm{Ga}$ impurity band levels, partially populated by holes, we use a positive (repulsive) 
charge background of $10^{18}$~cm$^{-3}$ in the (Ga,Mn)As layer leading to an upward shift of the potential profile in the well region, as shown in
Fig.~\ref{fig:ldos1}.  The local density of states of the ideal double-barrier structure (in absence of disorder) 
at zero bias and
$V = 0.1$~V,  corresponding to the dominant current peak of the IV curve (see Fig.~\ref{fig:IVkl}), is shown in
Fig.~\ref{fig:ldos1} and Fig.~\ref{fig:ldos2}, respectively.  The resonant levels
in the well are clearly visible and the solid line indicates the self-consistent potential profile of the structure.  At zero bias the valence band edge 
of (Ga,Mn)As is lifted relative to that of the contact GaAs layers by about 30~meV thus partially aligning the  Mn$_\mathrm{Ga}$ 
levels (indicated schematically by the bold (white) solid line) with the chemical potential which, in turn, lies about 10 meV above the valence band edge of the GaAs leads.  
Therefore, when a bias greater than about 10 mV is applied the  Mn$_\mathrm{Ga}$ 
levels can no longer be filled resonantly from the emitter side and, beyond about 30~mV, no longer from either emitter or the collector. 
 The latter situation is  shown schematically in  Fig.~\ref{fig:ldos2}.   This loss of holes from the  Mn$_\mathrm{Ga}$
 levels in the well region and the insufficient resupply of holes from the GaAs emitter into the resonant valence band levels 
lead to a breakdown of ferromagnetic order in the (Ga,Mn)As well under small bias, { \it i.e.} a zero bias anomaly.  


The transmission function and local density of states in the center of the Ga$_{1-x}$Mn$_x$As quantum well region is plotted for increasing degree of
disorder in Fig.~\ref{fig:T} and 
Fig.~\ref{fig:wdos}, respectively. Disorder leads to defect level in the band gap and a significant spectral broadening of the resonant levels associated with the lowest 
valence band resonances.  Note also the strong increase of transmission probability in the low--bias regime from disorder, opened by resonances for tunneling.

As a key difference compared to the bulk (Ga,Mn)As situation, the hole density inside a (Ga,Mn)As quantum well is established by forming a steady state situation 
with the external leads.   Since the Be doping level in the GaAs contacts is significantly lower than the Mn concentration this leads to a strongly reduced hole concentration 
(over bulk) in the (Ga,Mn)As wells under bias. 
Even under favorable bias conditions, in which resonant levels in the quantum well can be populated from the external leads, 
the quantum well hole density remains on the order of the hole density in the GaAs leads (in our case $\approx 10^{18}$~cm$^{-3}$), 
which is at least two orders of magnitude smaller than in bulk
(Ga,Mn)As. Therefore, in our simulations we find practically vanishing ferromagnetism (exchange splitting) for all bias values.  
In previous studies we have shown that for higher hole doping of the leads a bias-dependent ferromagnetic state with a maximum exchange splitting of the
order of tens of meV can be expected, being detectable by a significant spin polarization of the collector current.\cite{Ertler2011:PRB, Ertler2012:JCE}
In this earlier study we focused on the transport through the first HH-subband using a simpler effective mass band model.  
The picture that the quantum well hole density is determined by the coupling to the leads 
surely applies to the case when the impurity band
merges with the valence band, leading at most, to a broadening and shift of the valence band edge. 
In the presence of an isolated impurity band loosely bound holes may exist (at least at low voltages) in the confined impurity bands 
lying energetically  below the first valence 2D-subband (on the inverted energy scale). 

The main result of the paper is given in
Fig.~\ref{fig:IVkl}, which shows the IV-characteristics for increasing degree of disorder. 
Using a parallelized code
typically 240 configurations are used for each characteristic, which needs about 4 days of computation on a 12-node Opteron server for a single IV-curve.  
For increasing disorder the resonances in the IV-curve become more and more broadened and start to overlap until they are almost washed out.
Here a model which takes into account the HH-LH band mixing in the well is {\em essential} in order to see this effect.  If ignored,  the 
LH resonances would  dominate the current magnitude compared to the HH peaks and  regions of pronounced negative differential resistance would persist even for
unphysically high degrees of disorder. 
For the low doping regime of the leads (relative to the Mn concentration in the quantum well), as considered here, 
we find an almost vanishing ferromagnetism ($\Delta < 10^{-2}$~meV) in the well leading to a vanishing current spin polarization.
This results also suggest that, for a given (small) degree of disorder, enhancing the confinement, e.g. by using a thinner quantum well and/or higher
barriers, should lead to smaller overlap between the subbands, thus enhancing resonant--tunneling features  the IV-curve.   Alternative external pressure may be used to enhance the splitting between the lowest HH and LH subbands.

\section{Conclusions}\label{sec:sum}

In summary, we have shown that two factors can be relevant to understand the experimental findings of weak resonant--tunneling features and an absence of ferromagnetic order 
arising from thin (Ga,Mn)As quantum--well layers: 

(i) The presence of considerable disorder in thin layers of (Ga,Mn)As which conspires with 
HH-LH band mixing in the quantum well to efficiently weaken any  signatures of
resonant tunneling in the IV characteristics.

(ii) The observed (almost) vanishing of ferromagnetic order  in the (Ga,Mn)As quantum well can be understood by
finding orders of magnitude lower hole 
densities
in the well as compared to the bulk (Ga,Mn)As case of equal Mn concentration.   

We thus find the original interpretation of observing quantum size effects in a (Ga,Mn)As quantum well by the group of Tanaka 
\cite{Ohya2011:NP, Ohya2007:PRB, Ohya2010:PRL} 
to be plausible and consistent with our numerical results.

\section{Acknowledgment}

We acknowledge helpful discussions with Prof. Masaaki Tanaka. This work has been supported by the Austrian Science Foundation under FWF project P21289-N16.


\begin{thebibliography}{28}
\expandafter\ifx\csname natexlab\endcsname\relax\def\natexlab#1{#1}\fi
\expandafter\ifx\csname bibnamefont\endcsname\relax
  \def\bibnamefont#1{#1}\fi
\expandafter\ifx\csname bibfnamefont\endcsname\relax
  \def\bibfnamefont#1{#1}\fi
\expandafter\ifx\csname citenamefont\endcsname\relax
  \def\citenamefont#1{#1}\fi
\expandafter\ifx\csname url\endcsname\relax
  \def\url#1{\texttt{#1}}\fi
\expandafter\ifx\csname urlprefix\endcsname\relax\def\urlprefix{URL }\fi
\providecommand{\bibinfo}[2]{#2}
\providecommand{\eprint}[2][]{\url{#2}}

\bibitem[{\citenamefont{Jungwirth et~al.}(2006)\citenamefont{Jungwirth, Sinova,
  Ma\v{s}ek, Ku\v{c}era, and MacDonald}}]{Jungwirth2006:RMP}
\bibinfo{author}{\bibfnamefont{T.}~\bibnamefont{Jungwirth}},
  \bibinfo{author}{\bibfnamefont{J.}~\bibnamefont{Sinova}},
  \bibinfo{author}{\bibfnamefont{J.}~\bibnamefont{Ma\v{s}ek}},
  \bibinfo{author}{\bibfnamefont{J.}~\bibnamefont{Ku\v{c}era}},
  \bibnamefont{and} \bibinfo{author}{\bibfnamefont{A.~H.}
  \bibnamefont{MacDonald}}, \bibinfo{journal}{Rev. Mod. Phys.}
  \textbf{\bibinfo{volume}{78}}, \bibinfo{pages}{809} (\bibinfo{year}{2006}).

\bibitem[{\citenamefont{Burch et~al.}(2008)\citenamefont{Burch, Awschalom, and
  Basov}}]{Burch2008:JMMM}
\bibinfo{author}{\bibfnamefont{K.~S.} \bibnamefont{Burch}},
  \bibinfo{author}{\bibfnamefont{D.~D.} \bibnamefont{Awschalom}},
  \bibnamefont{and} \bibinfo{author}{\bibfnamefont{D.~N.} \bibnamefont{Basov}},
  \bibinfo{journal}{J. Magn. Magn. Mat.} \textbf{\bibinfo{volume}{320}},
  \bibinfo{pages}{3207} (\bibinfo{year}{2008}).

\bibitem[{\citenamefont{Schneider et~al.}(1987)\citenamefont{Schneider,
  Kaufmann, Wilkening, Baeumler, and K{\"o}hl}}]{Schneider1987:PRL}
\bibinfo{author}{\bibfnamefont{J.}~\bibnamefont{Schneider}},
  \bibinfo{author}{\bibfnamefont{U.}~\bibnamefont{Kaufmann}},
  \bibinfo{author}{\bibfnamefont{W.}~\bibnamefont{Wilkening}},
  \bibinfo{author}{\bibfnamefont{M.}~\bibnamefont{Baeumler}}, \bibnamefont{and}
  \bibinfo{author}{\bibfnamefont{F.}~\bibnamefont{K{\"o}hl}},
  \bibinfo{journal}{Phys. Rev. Lett.} \textbf{\bibinfo{volume}{59}},
  \bibinfo{pages}{240} (\bibinfo{year}{1987}).

\bibitem[{\citenamefont{Jungwirth et~al.}(2007)\citenamefont{Jungwirth, Sinova,
  MacDonald, Gallagher, Nov\'ak, Edmonds, Rushforth, Campion, Foxon, Eaves
  et~al.}}]{Jungwirth2007:PRB}
\bibinfo{author}{\bibfnamefont{T.}~\bibnamefont{Jungwirth}},
  \bibinfo{author}{\bibfnamefont{J.}~\bibnamefont{Sinova}},
  \bibinfo{author}{\bibfnamefont{A.~H.} \bibnamefont{MacDonald}},
  \bibinfo{author}{\bibfnamefont{B.~L.} \bibnamefont{Gallagher}},
  \bibinfo{author}{\bibfnamefont{V.}~\bibnamefont{Nov\'ak}},
  \bibinfo{author}{\bibfnamefont{K.~W.} \bibnamefont{Edmonds}},
  \bibinfo{author}{\bibfnamefont{A.~W.} \bibnamefont{Rushforth}},
  \bibinfo{author}{\bibfnamefont{R.~P.} \bibnamefont{Campion}},
  \bibinfo{author}{\bibfnamefont{C.~T.} \bibnamefont{Foxon}},
  \bibinfo{author}{\bibfnamefont{L.}~\bibnamefont{Eaves}},
  \bibnamefont{et~al.}, \bibinfo{journal}{Phys. Rev. B}
  \textbf{\bibinfo{volume}{76}}, \bibinfo{pages}{125206}
  (\bibinfo{year}{2007}).

\bibitem[{\citenamefont{Richardella et~al.}(2010)\citenamefont{Richardella,
  Roushan, Mack, Zhou, Huse, Awschalom, and Yazdani}}]{Richardella2010:S}
\bibinfo{author}{\bibfnamefont{A.}~\bibnamefont{Richardella}},
  \bibinfo{author}{\bibfnamefont{P.}~\bibnamefont{Roushan}},
  \bibinfo{author}{\bibfnamefont{S.}~\bibnamefont{Mack}},
  \bibinfo{author}{\bibfnamefont{B.}~\bibnamefont{Zhou}},
  \bibinfo{author}{\bibfnamefont{D.~A.} \bibnamefont{Huse}},
  \bibinfo{author}{\bibfnamefont{D.~D.} \bibnamefont{Awschalom}},
  \bibnamefont{and} \bibinfo{author}{\bibfnamefont{A.}~\bibnamefont{Yazdani}},
  \bibinfo{journal}{Science} \textbf{\bibinfo{volume}{327}},
  \bibinfo{pages}{665} (\bibinfo{year}{2010}).

\bibitem[{\citenamefont{Burch et~al.}(2006)\citenamefont{Burch, Shrekenhamer,
  Singley, Stephens, Sheu, Kawakami, Schiffer, Samarth, Awschalom, and
  Basov}}]{Burch2006:PRL}
\bibinfo{author}{\bibfnamefont{K.~S.} \bibnamefont{Burch}},
  \bibinfo{author}{\bibfnamefont{D.~B.} \bibnamefont{Shrekenhamer}},
  \bibinfo{author}{\bibfnamefont{E.~J.} \bibnamefont{Singley}},
  \bibinfo{author}{\bibfnamefont{J.}~\bibnamefont{Stephens}},
  \bibinfo{author}{\bibfnamefont{B.~L.} \bibnamefont{Sheu}},
  \bibinfo{author}{\bibfnamefont{R.~K.} \bibnamefont{Kawakami}},
  \bibinfo{author}{\bibfnamefont{P.}~\bibnamefont{Schiffer}},
  \bibinfo{author}{\bibfnamefont{N.}~\bibnamefont{Samarth}},
  \bibinfo{author}{\bibfnamefont{D.~D.} \bibnamefont{Awschalom}},
  \bibnamefont{and} \bibinfo{author}{\bibfnamefont{D.~N.} \bibnamefont{Basov}},
  \bibinfo{journal}{Phys. Rev. Lett.} \textbf{\bibinfo{volume}{97}},
  \bibinfo{pages}{87208} (\bibinfo{year}{2006}).

\bibitem[{\citenamefont{Ohya et~al.}(2011)\citenamefont{Ohya, Takata, and
  Tanaka}}]{Ohya2011:NP}
\bibinfo{author}{\bibfnamefont{S.}~\bibnamefont{Ohya}},
  \bibinfo{author}{\bibfnamefont{K.}~\bibnamefont{Takata}}, \bibnamefont{and}
  \bibinfo{author}{\bibfnamefont{M.}~\bibnamefont{Tanaka}},
  \bibinfo{journal}{Nature Physics} \textbf{\bibinfo{volume}{7}},
  \bibinfo{pages}{342} (\bibinfo{year}{2011}).

\bibitem[{\citenamefont{Ohya et~al.}(2007)\citenamefont{Ohya, Hai, Mizuno, and
  Tanaka}}]{Ohya2007:PRB}
\bibinfo{author}{\bibfnamefont{S.}~\bibnamefont{Ohya}},
  \bibinfo{author}{\bibfnamefont{P.~N.} \bibnamefont{Hai}},
  \bibinfo{author}{\bibfnamefont{Y.}~\bibnamefont{Mizuno}}, \bibnamefont{and}
  \bibinfo{author}{\bibfnamefont{M.}~\bibnamefont{Tanaka}},
  \bibinfo{journal}{Phys. Rev. B} \textbf{\bibinfo{volume}{75}},
  \bibinfo{pages}{155328} (\bibinfo{year}{2007}).

\bibitem[{\citenamefont{Ohya et~al.}(2010)\citenamefont{Ohya, Muneta, Hai, and
  Tanaka}}]{Ohya2010:PRL}
\bibinfo{author}{\bibfnamefont{S.}~\bibnamefont{Ohya}},
  \bibinfo{author}{\bibfnamefont{I.}~\bibnamefont{Muneta}},
  \bibinfo{author}{\bibfnamefont{P.~N.} \bibnamefont{Hai}}, \bibnamefont{and}
  \bibinfo{author}{\bibfnamefont{M.}~\bibnamefont{Tanaka}},
  \bibinfo{journal}{Phys. Rev. Lett.} \textbf{\bibinfo{volume}{104}},
  \bibinfo{pages}{167204} (\bibinfo{year}{2010}).

\bibitem[{\citenamefont{{Dietl} and {Sztenkiel}}(2011)}]{Dietl:arXiv1102.3267D}
\bibinfo{author}{\bibfnamefont{T.}~\bibnamefont{{Dietl}}} \bibnamefont{and}
  \bibinfo{author}{\bibfnamefont{D.}~\bibnamefont{{Sztenkiel}}},
  \bibinfo{journal}{ArXiv e-prints}  (\bibinfo{year}{2011}),
  \eprint{1102.3267}.

\bibitem[{\citenamefont{{Ohya} et~al.}(2011)\citenamefont{{Ohya}, {Takata},
  {Muneta}, {Hai}, and {Tanaka}}}]{Ohya:arXiv1102.4459O}
\bibinfo{author}{\bibfnamefont{S.}~\bibnamefont{{Ohya}}},
  \bibinfo{author}{\bibfnamefont{K.}~\bibnamefont{{Takata}}},
  \bibinfo{author}{\bibfnamefont{I.}~\bibnamefont{{Muneta}}},
  \bibinfo{author}{\bibfnamefont{P.~N.} \bibnamefont{{Hai}}}, \bibnamefont{and}
  \bibinfo{author}{\bibfnamefont{M.}~\bibnamefont{{Tanaka}}},
  \bibinfo{journal}{ArXiv e-prints}  (\bibinfo{year}{2011}),
  \eprint{1102.4459}.

\bibitem[{\citenamefont{Madelung}(1978)}]{Madelung:1978}
\bibinfo{author}{\bibfnamefont{O.}~\bibnamefont{Madelung}},
  \emph{\bibinfo{title}{Introduction to Solid-State Theory}}
  (\bibinfo{publisher}{Springer, Berlin}, \bibinfo{year}{1978}).

\bibitem[{\citenamefont{Likovich et~al.}(2009)\citenamefont{Likovich, Russell,
  Yi, Narayanamurti, Ku, Zhu, and Samarth}}]{Likovich2009:PRB}
\bibinfo{author}{\bibfnamefont{E.}~\bibnamefont{Likovich}},
  \bibinfo{author}{\bibfnamefont{K.}~\bibnamefont{Russell}},
  \bibinfo{author}{\bibfnamefont{W.}~\bibnamefont{Yi}},
  \bibinfo{author}{\bibfnamefont{V.}~\bibnamefont{Narayanamurti}},
  \bibinfo{author}{\bibfnamefont{K.-C.} \bibnamefont{Ku}},
  \bibinfo{author}{\bibfnamefont{M.}~\bibnamefont{Zhu}}, \bibnamefont{and}
  \bibinfo{author}{\bibfnamefont{N.}~\bibnamefont{Samarth}},
  \bibinfo{journal}{Phys. Rev. B} \textbf{\bibinfo{volume}{80}},
  \bibinfo{pages}{201307(R)} (\bibinfo{year}{2009}).

\bibitem[{\citenamefont{Miyazaki et~al.}(1987)\citenamefont{Miyazaki, Ihara,
  and Hirose}}]{Miyazaki1987:PRL}
\bibinfo{author}{\bibfnamefont{S.}~\bibnamefont{Miyazaki}},
  \bibinfo{author}{\bibfnamefont{Y.}~\bibnamefont{Ihara}}, \bibnamefont{and}
  \bibinfo{author}{\bibfnamefont{M.}~\bibnamefont{Hirose}},
  \bibinfo{journal}{Phys. Rev. Lett.} \textbf{\bibinfo{volume}{59}},
  \bibinfo{pages}{125} (\bibinfo{year}{1987}).

\bibitem[{\citenamefont{Li and P{\"o}tz}(1993)}]{Li1993:PRB}
\bibinfo{author}{\bibfnamefont{Z.~Q.} \bibnamefont{Li}} \bibnamefont{and}
  \bibinfo{author}{\bibfnamefont{W.}~\bibnamefont{P{\"o}tz}},
  \bibinfo{journal}{Phys. Rev. B} \textbf{\bibinfo{volume}{47}},
  \bibinfo{pages}{6509} (\bibinfo{year}{1993}).

\bibitem[{\citenamefont{{Van Esch} et~al.}(1997)\citenamefont{{Van Esch}, {Van
  Bockstal}, {De Boeck}, Verbanck, {van Steenbergen}, Wellmann, Grietens,
  Bogaerts, Herlach, and Borghs}}]{VanEsch1997:PRB}
\bibinfo{author}{\bibfnamefont{A.}~\bibnamefont{{Van Esch}}},
  \bibinfo{author}{\bibfnamefont{L.}~\bibnamefont{{Van Bockstal}}},
  \bibinfo{author}{\bibfnamefont{J.}~\bibnamefont{{De Boeck}}},
  \bibinfo{author}{\bibfnamefont{G.}~\bibnamefont{Verbanck}},
  \bibinfo{author}{\bibfnamefont{A.~S.} \bibnamefont{{van Steenbergen}}},
  \bibinfo{author}{\bibfnamefont{P.~J.} \bibnamefont{Wellmann}},
  \bibinfo{author}{\bibfnamefont{B.}~\bibnamefont{Grietens}},
  \bibinfo{author}{\bibfnamefont{R.}~\bibnamefont{Bogaerts}},
  \bibinfo{author}{\bibfnamefont{F.}~\bibnamefont{Herlach}}, \bibnamefont{and}
  \bibinfo{author}{\bibfnamefont{G.}~\bibnamefont{Borghs}},
  \bibinfo{journal}{Phys. Rev. B} \textbf{\bibinfo{volume}{56}},
  \bibinfo{pages}{13103} (\bibinfo{year}{1997}).

\bibitem[{\citenamefont{{Das Sarma} et~al.}(2003)\citenamefont{{Das Sarma},
  Hwang, and Kaminski}}]{DasSarma2003:PRB}
\bibinfo{author}{\bibfnamefont{S.}~\bibnamefont{{Das Sarma}}},
  \bibinfo{author}{\bibfnamefont{E.~H.} \bibnamefont{Hwang}}, \bibnamefont{and}
  \bibinfo{author}{\bibfnamefont{A.}~\bibnamefont{Kaminski}},
  \bibinfo{journal}{Phys. Rev. B} \textbf{\bibinfo{volume}{67}},
  \bibinfo{pages}{155201} (\bibinfo{year}{2003}).

\bibitem[{\citenamefont{Ertler and P\"otz}(2011)}]{Ertler2011:PRB}
\bibinfo{author}{\bibfnamefont{C.}~\bibnamefont{Ertler}} \bibnamefont{and}
  \bibinfo{author}{\bibfnamefont{W.}~\bibnamefont{P\"otz}},
  \bibinfo{journal}{Phys. Rev. B} \textbf{\bibinfo{volume}{84}},
  \bibinfo{pages}{165309} (\bibinfo{year}{2011}).

\bibitem[{\citenamefont{Ertler and P\"otz}(2012)}]{Ertler2012:JCE}
\bibinfo{author}{\bibfnamefont{C.}~\bibnamefont{Ertler}} \bibnamefont{and}
  \bibinfo{author}{\bibfnamefont{W.}~\bibnamefont{P\"otz}},
  \bibinfo{journal}{Journal of Computational Electronics} pp.
  \bibinfo{pages}{1--7} (\bibinfo{year}{2012}), ISSN \bibinfo{issn}{1569-8025}.

\bibitem[{\citenamefont{Luttinger and Kohn}(1955)}]{Luttinger1955:PR}
\bibinfo{author}{\bibfnamefont{J.~M.} \bibnamefont{Luttinger}}
  \bibnamefont{and} \bibinfo{author}{\bibfnamefont{W.}~\bibnamefont{Kohn}},
  \bibinfo{journal}{Phys. Rev.} \textbf{\bibinfo{volume}{97}},
  \bibinfo{pages}{869} (\bibinfo{year}{1955}).

\bibitem[{\citenamefont{Chao and Chuang}(1991)}]{Chao1991:PRB}
\bibinfo{author}{\bibfnamefont{C.~Y.-P.} \bibnamefont{Chao}} \bibnamefont{and}
  \bibinfo{author}{\bibfnamefont{S.~L.} \bibnamefont{Chuang}},
  \bibinfo{journal}{Phys. Rev. B} \textbf{\bibinfo{volume}{43}},
  \bibinfo{pages}{7027} (\bibinfo{year}{1991}).

\bibitem[{\citenamefont{Dietl et~al.}(1997)\citenamefont{Dietl, Haury, and
  d'Aubign\'e}}]{Dietl1997:PRB}
\bibinfo{author}{\bibfnamefont{T.}~\bibnamefont{Dietl}},
  \bibinfo{author}{\bibfnamefont{A.}~\bibnamefont{Haury}}, \bibnamefont{and}
  \bibinfo{author}{\bibfnamefont{Y.~M.} \bibnamefont{d'Aubign\'e}},
  \bibinfo{journal}{Phys. Rev. B} \textbf{\bibinfo{volume}{55}},
  \bibinfo{pages}{R3347} (\bibinfo{year}{1997}).

\bibitem[{\citenamefont{Jungwirth et~al.}(1999)\citenamefont{Jungwirth,
  Atkinson, Lee, and MacDonald}}]{Jungwirth1999:PRB}
\bibinfo{author}{\bibfnamefont{T.}~\bibnamefont{Jungwirth}},
  \bibinfo{author}{\bibfnamefont{W.~A.} \bibnamefont{Atkinson}},
  \bibinfo{author}{\bibfnamefont{B.~H.} \bibnamefont{Lee}}, \bibnamefont{and}
  \bibinfo{author}{\bibfnamefont{A.~H.} \bibnamefont{MacDonald}},
  \bibinfo{journal}{Phys. Rev. B} \textbf{\bibinfo{volume}{59}},
  \bibinfo{pages}{9818} (\bibinfo{year}{1999}).

\bibitem[{\citenamefont{Fabian et~al.}(2007)\citenamefont{Fabian,
  Matos-Abiague, Ertler, Stano, and {\v{Z}uti\'c}}}]{Fabian2007:APS}
\bibinfo{author}{\bibfnamefont{J.}~\bibnamefont{Fabian}},
  \bibinfo{author}{\bibfnamefont{A.}~\bibnamefont{Matos-Abiague}},
  \bibinfo{author}{\bibfnamefont{C.}~\bibnamefont{Ertler}},
  \bibinfo{author}{\bibfnamefont{P.}~\bibnamefont{Stano}}, \bibnamefont{and}
  \bibinfo{author}{\bibfnamefont{I.}~\bibnamefont{{\v{Z}uti\'c}}},
  \bibinfo{journal}{Acta Physica Slovaca} \textbf{\bibinfo{volume}{57}},
  \bibinfo{pages}{565} (\bibinfo{year}{2007}).

\bibitem[{\citenamefont{Datta}(1995)}]{Datta:1995}
\bibinfo{author}{\bibfnamefont{S.}~\bibnamefont{Datta}},
  \emph{\bibinfo{title}{Electronic Transport in Mesoscopic Systems}}
  (\bibinfo{publisher}{Cambridge University Press, Cambridge, England},
  \bibinfo{year}{1995}).

\bibitem[{\citenamefont{L{\'o}pez-Sancho
  et~al.}(1985)\citenamefont{L{\'o}pez-Sancho, L{\'o}pez-Sancho, and
  Rubio}}]{Sancho1985:JPF}
\bibinfo{author}{\bibfnamefont{M.~P.} \bibnamefont{L{\'o}pez-Sancho}},
  \bibinfo{author}{\bibfnamefont{J.~M.} \bibnamefont{L{\'o}pez-Sancho}},
  \bibnamefont{and} \bibinfo{author}{\bibfnamefont{J.}~\bibnamefont{Rubio}},
  \bibinfo{journal}{J. Phys. F: Met. Phys.} \textbf{\bibinfo{volume}{15}},
  \bibinfo{pages}{851} (\bibinfo{year}{1985}).

\bibitem[{\citenamefont{Kaxiras}(2003)}]{Kaxiras:2003}
\bibinfo{author}{\bibfnamefont{E.}~\bibnamefont{Kaxiras}},
  \emph{\bibinfo{title}{Atomic and Electronic Structure of Solids}}
  (\bibinfo{publisher}{Cambridge University Press, Cambridge},
  \bibinfo{year}{2003}).

\bibitem[{\citenamefont{Ashcroft and Mermin}(1976)}]{Ashcroft:1976}
\bibinfo{author}{\bibfnamefont{N.~W.} \bibnamefont{Ashcroft}} \bibnamefont{and}
  \bibinfo{author}{\bibfnamefont{N.~D.} \bibnamefont{Mermin}},
  \emph{\bibinfo{title}{Solid State Physics}} (\bibinfo{publisher}{Saunders,
  Philadelphia}, \bibinfo{year}{1976}).

\end{thebibliography}

\end{document}